\documentclass[namedreferences]{SolarPhysics}
\usepackage[optionalrh]{spr-sola-addons} 
\usepackage[pdftex]{graphicx} 
\usepackage{color}           
\usepackage{url}             

\begin{document}

\begin{article}

\begin{opening}

\title{Radio Bursts Associated with Flare and Ejecta in the 13 July 
2004 Event}

\author{S.~\surname{Pohjolainen}\sep
        K.~\surname{Hori}\sep
        T.~\surname{Sakurai}
       }
\runningauthor{Pohjolainen, Hori, and Sakurai}
\runningtitle{Radio Bursts Associated with Flare and Ejecta}

\institute{S.~\surname{Pohjolainen}\\
           Department of Physics and Astronomy, University of Turku, 
           Tuorla Observatory, Piikki\"o, Finland email: \url{silpoh@utu.fi}\\
           K.~\surname{Hori}, T.~\surname{Sakurai}\\
           National Astronomical Observatory of Japan, 
           Mitaka, Tokyo, Japan 
           email: \url{kuniko.hori@thomsonreuters.com}, 
           \url{sakurai@solar.mtk.nao.ac.jp}}

\begin{abstract}
We investigate coronal transients associated with a GOES M6.7 class 
flare and a coronal mass ejection (CME) on 13 July 2004. 
During the rising phase of the flare, a filament eruption, loop 
expansion, a Moreton wave, and an ejecta were observed. An EIT wave 
was detected later on. The main features in the radio dynamic 
spectrum were a frequency-drifting continuum and two type II bursts. 
Our analysis shows that if the first type II burst was formed in
the low corona, the burst heights and speed are close to the projected 
distances and speed of the Moreton wave (a chromospheric shock wave 
signature). The frequency-drifting radio continuum, starting 
above 1 GHz, was formed almost two minutes prior to any shock features 
becoming visible, and a fast-expanding piston (visible as the continuum) 
could have launched another shock wave. A possible scenario is that a 
flare blast overtook the earlier transient, and ignited the first type 
II burst. The second type II burst may have been formed by the same 
shock, but only if the shock was propagating at a constant speed. 
This interpretation also requires that the shock-producing regions 
were located at different parts of the propagating structure, or that 
the shock was passing through regions with highly different atmospheric 
densities. This complex event, with a multitude of radio features 
and transients at other wavelengths, presents evidence for both 
blast-wave-related and CME-related radio emissions.
\end{abstract}

\end{opening}


\section{Introduction}

Decimetric--metric radio emission is usually associated with plasma waves 
excited by accelerated electrons, where the observed emission frequency 
is directly related to the plasma density in the affected region
\cite{melrose80}.  
Fast-drifting structures in radio dynamic spectra, such as type III bursts,  
are normally attributed to electron beams propagating outward in the solar 
atmosphere. The type III burst-associated acceleration usually happens
in connection with flaring processes, by for example, reconnecting field
lines. A slower drift and a broader instantaneous emission band in a 
radio burst can indicate a moving shock front that accelerates electrons. 
Some of these bursts can be classified as type II bursts, in particular 
when both fundamental and second harmonic emission bands are visible 
({\it e.g.}, Nelson and Melrose, \citeyear{nelson85}; Mann, Classen, 
and Aurass, \citeyear{mann95}; Cairns {\it et al.}, \citeyear{cairns03}). 
Although type II bursts indicate the existence of shock waves, their 
origin is not always so clear. Shocks can basically be flare related or 
coronal mass ejection (CME) related, but coronal (metric) type II bursts 
do not necessarily have the same origin as the longer wavelength 
(interplanetary) type II bursts (Mancuso and Raymond, \citeyear{mancuso04}; 
Cane and Erickson, \citeyear{cane05}; Lin, Mancuso, and Vourlidas, 
\citeyear{lin06}, and references therein). 

Theoretically, shocks can be freely propagating (blast) waves, piston-driven 
shocks formed either at the nose or the flanks of an expanding body, or 
bow shocks ahead of a projectile \cite{warmuth07}. A freely propagating 
shock can also develop if a piston stops or slows down considerably. 
One characteristic of piston-driven shocks is that the speed of the 
propagating shock is different from the speed of the driver, and the 
shock speed should be higher. For bow shocks ahead of a projectile, the 
speeds should be roughly the same, see, for example, \inlinecite{vrsnak05}.  
   
Frequency-drifting continuum emissions at centimeter to meter wavelengths, 
also known as type IV bursts, have been associated with outward moving 
plasmoids, see review on solar radio emission in, for example, 
\inlinecite{dulk85}. At decimetric wavelengths, the usually smooth type 
IV continuum may include pulsating structures \cite{kundu65}. 
In some cases decimetric pulsating structures can be taken as precursors 
to type II bursts \cite{klassen99}, and identified as a signature of 
reconnection processes above expanding soft X-ray loops \cite{klassen03}.

This study presents a flare--CME event on 13 July 2004, during which 
a rich variety of radio bursts were observed from centimeter to meter 
wavelengths. In this paper, we concentrate on the type II and type IV
bursts observed at decimeter--meter wavelengths. During this event, 
quasi-periodic pulsations were observed in microwaves and decimeter 
waves, but they will be discussed in another paper.  
We compare the timing, heights, and structural evolution of the observed 
features and discuss conditions that are needed for the radio emissions 
to appear. 

\section{Observations}

Full-disk EUV images at 195 \AA \  during the 13 July 2004 event were 
provided by the SOHO EUV Imaging Telescope (EIT) instrument \cite{boudin95}, 
and high resolution, limited field of view images at 171 \AA \ by the 
Transition Region and Coronal Explorer (TRACE) satellite \cite{handy}. 
Big Bear Solar Observatory (BBSO; part of the global high-resolution 
H$\alpha$ network at 656.3 nm wavelength) observed the event in 
H$\alpha$ with a cadence of 1\,--\,3 images per minute.  
White-light coronagraph images were obtained from the SOHO Large Angle 
Spectrometric Coronagraph (LASCO) instrument \cite{brueckner95}, 
and the Reuven Ramaty High Energy Solar Spectroscopic Imager (RHESSI)
provided hard X-ray data \cite{lin02}. 
Coronal mass ejections are listed in the LASCO CME Catalog 
at \texttt{http://cdaw.gsfc.nasa.gov/CME\_list/}. 

Nobeyama radio polarimeters (NoRP) observed radio flux densities 
at seven single frequencies between 1 and 80 GHz \cite{nakajima85}. 
Simultaneous radio imaging was obtained at 17 and 34 GHz with the 
Nobeyama Radioheliograph, NoRH \cite{nakajima94}. At decimetric-metric 
wavelengths dynamic spectra were recorded by the HiRAS radiospectrograph, 
which covers a frequency range from 25 MHz to 2.5 GHz, and with the Radio 
Solar Telescope Network (RSTN) instruments operated by the U.S. Air Force 
at four different sites, observing at 25--180 MHz. For comparison, 
NoRP and HiRAS have two overlapping frequencies, at 1 and 2 GHz.

\subsection{Flare,  Ejecta and EIT Wave}

On 13 July 2004 a GOES M6.7 class flare was observed in the NOAA active 
region 10646, located at N14\,W45. The GOES soft X-ray flux started to 
rise at 00:08 UT, and there were two impulsive rises after that, at 
00:11:40 and 00:13:05 UT. Flux peak was recorded at 00:17 UT.

The first signs of a filament eruption were visible in the TRACE EUV 
image at 00:12:45 UT. At about the same time, quasi-periodic pulsations 
in hard X-rays and microwaves started. H$\alpha$ flare ribbons 
appeared in the images at 00:14:25 UT.

\begin{figure}
\includegraphics[width=12cm,angle=0]{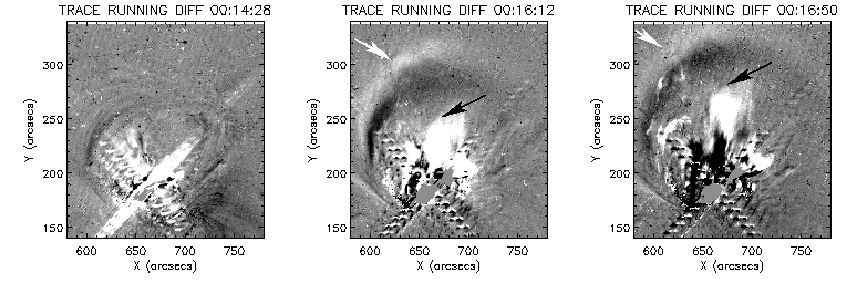}
\includegraphics[width=12cm,angle=0]{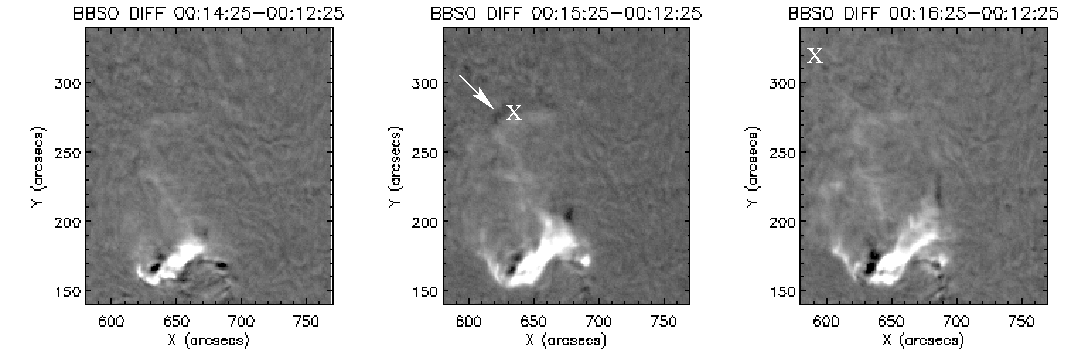}
  \caption{Top: TRACE difference images showing a rising loop-like 
    structure (white arrows; projected speed 235\,--\,330 km s$^{-1}$) 
    above the active region at N14W45, and a following ejecta (black arrows; 
    projected speed of 660 km s$^{-1}$). 
    Bottom: BBSO H$\alpha$ difference images of the same region. 
    A chromospheric disturbance is observed near the location 
    of the rising EUV loop-like structure (white arrow). Crosses mark 
    the locations of the H$\alpha$ Moreton wave front (which are too faint 
    to be observed in these images, see text for details).   
    }
\label{fig:fig1}
\end{figure}

A large EUV loop structure was observed over the active region before the 
flare started, and this structure began to rise at about 00:14 UT. 
The arc-shaped front is indicated with white arrows in the TRACE difference 
images in Figure \ref{fig:fig1} (top row). The structure moved outside 
the TRACE field of view at about 00:18 UT. The projected plane-of-the-sky 
rise velocity of the front is estimated to be between 235 and 330 
km s$^{-1}$. Some caution is needed when interpreting propagating 
loop-like features, as some of them can be shock waves driven by erupting 
filaments \cite{pomoell08}, but in this case the pre-event EUV loop and 
the slow rise velocity suggest that it was a rising loop. 

At about 00:15 UT, below the rising loop-like structure, a bright EUV 
ejecta started to move upward. The ejecta front is indicated with black 
arrows in Figure \ref{fig:fig1} (top row). The projected rise speed is 
estimated to be about 660 km s$^{-1}$.  

H$\alpha$ difference images show a disturbance at the chromospheric level 
at about 00:15 UT (at a location indicated with a white arrow in Figure 
\ref{fig:fig1}, bottom row). This feature is very near the location 
of the EUV loop-like structure, and also near the front of a faint 
H$\alpha$ Moreton wave (Figure \ref{fig:fig1}, with locations indicated with 
crosses); see details and a movie in \inlinecite{grechnev08}. 
The H$\alpha$ wave was propagating toward North--Northeast from the 
active region and the projected speed estimate is about 700 km s$^{-1}$.
H$\alpha$ (Moreton) waves are generally thought to be direct signatures 
of propagating blast wave shocks in the chromosphere \cite{uchida74}. 

\begin{figure}
\centering
\includegraphics[width=12cm,angle=0]{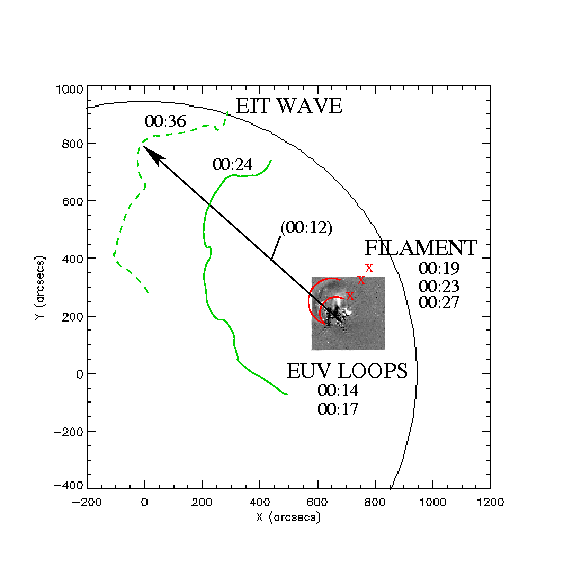}
  \caption{EIT wave locations at 00:24 and 00:36 UT, with an estimate
           of the wave location at 00:12 UT if it had moved at a
           constant speed. Note that the EIT image at 00:12 UT does not show
           any signatures of an EIT wave. The expanding and rising 
           loop-like structure observed by TRACE in EUV is shown in the 
           insert image (outlined in black solid lines) and Xs mark 
           the path of the erupting H$\alpha$ filament observed by BBSO.}
\label{fig:fig2}
\end{figure}

Later on, the erupting H$\alpha$ filament was observed to move toward 
the Northwest. The filament trajectory, at three different times between
00:19 and 00:27 UT traced from the BBSO H$\alpha$ observations, is 
marked in \mbox{Figure \ref{fig:fig2}}.  

In the full-disk EUV images (SOHO EIT) brightenings and dimmed regions 
can be observed from 00:24 UT onward. The brightenings in the EIT 
difference images are the so-called EIT-waves \cite{thompson99}, and we 
can estimate the wave front distances from the active region at two 
different times. The estimated (projected) speed of the front, measured 
from the running difference images at 00:24 and 00:36 UT, is approximately 
300 km s$^{-1}$. Figure \ref{fig:fig2} shows the outlines of the farthermost 
fronts toward the Northeast at 00:24 and 00:36 UT, and also the 
backward-extrapolated front distance at 00:12 UT, if the wave speed had 
remained constant. However, in the EIT image at 00:12 UT no EIT wave can 
be detected. Either the EIT wave had formed some distance away from the 
active region or the wave had decelerated after formation. The rising EUV 
loop front locations at 00:14 and 00:17 UT, along with the erupting filament 
trajectory, are indicated in Figure \ref{fig:fig2} for comparison.

\begin{figure}
\includegraphics[width=12cm,angle=0]{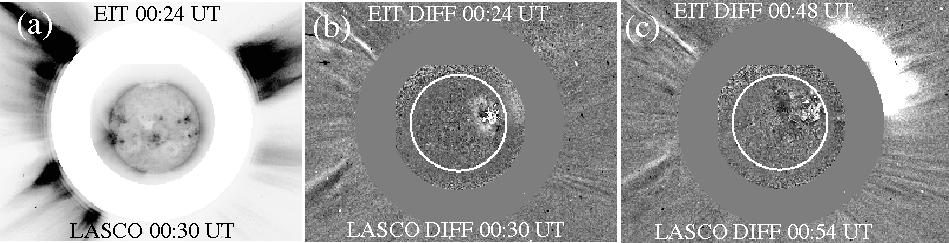}
  \caption{LASCO C2 image at (a) 00:30 UT showing a strong streamer 
           region in the Northwest direction. 
           LASCO C2 running difference images at (b) 00:30 UT and 
           (c) 00:54 UT, with the EIT difference images inside the 
           occulted disk, showing the early evolution of the CME. 
           (Images courtesy of the LASCO CME Catalog)}
\label{fig:fig3}
\end{figure}

\subsection{Coronal Mass Ejection}

A CME was first detected in white-light in the LASCO C2 difference image 
at 00:54 UT, over the solar northwestern limb (Figure \ref{fig:fig3}) 
at a heliocentric height of 3.07 $R_{\odot}$. In the previous LASCO image 
at 00:30 UT the CME front is not yet visible, but the SOHO EIT running 
difference image at 00:24 UT shows brightenings on the disk and above 
the limb in the same direction as the later-observed CME. In the 
pre-eruption images an intense streamer region is visible 
over the northwestern limb (Figure \ref{fig:fig3}, left panel).
  
The projected plane-of-the-sky speed of the CME was approximately 
\mbox{450 km s$^{-1}$}, measured from the LASCO images at 00:54 and 01:31 UT. 
Later on, the CME decelerated to a speed of about 400 km s$^{-1}$ 
(LASCO CME Catalog, linear fit to all later CME observations and 
second order fit to all data). The CME Catalog also lists another 
event during the same period, with a plane-of-the sky speed of 
\mbox{600 km s$^{-1}$}, but from the LASCO images this leading edge is 
hard to identify.

\subsection{Radio Emission}

\begin{figure}[!h]
\includegraphics[width=12cm,angle=0]{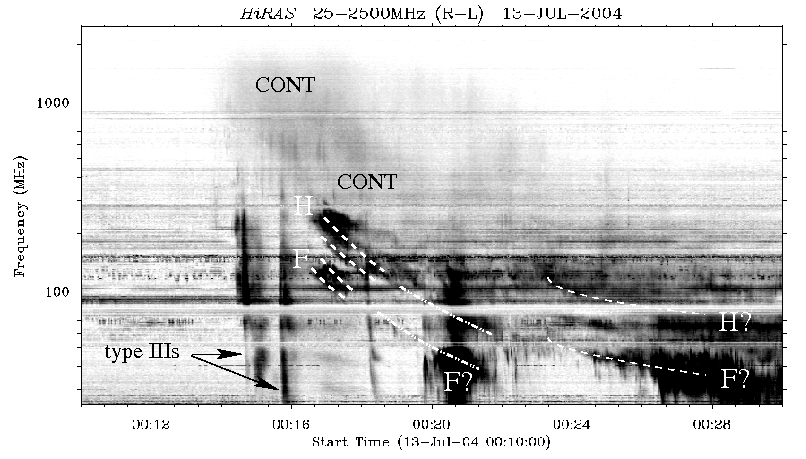}
  \caption{HiRAS dynamic radio spectrum at 00:10\,--\,00:30 UT at 
           25 MHz\,--\,2.5 GHz. At least two type II burst lanes, with 
           emission at the fundamental (F) and second harmonic (H) frequencies, 
           can be found in the spectrum. The emission lanes of the 
           first type II burst are also band-split. Frequency-drifting 
           continuum emission (labeled CONT) is visible below 
           \mbox{$\approx$ 2 GHz}. 
}  
\label{fig:fig4}
\end{figure}

HiRAS radio dynamic spectrum shows that radio emission at 
decimetric--metric wavelengths consist of several emission features 
(Figure \ref{fig:fig4}). First, it shows a faint, frequency-drifting 
continuum starting above 1 GHz. At 1 GHz, the continuum flux starts 
to rise at about 00:13:20 UT. 
Metric type III bursts that continue to decameter --hectometer wavelengths 
({\it Wind} WAVES observations) are observed soon after the appearance 
of the drifting continuum. This indicates that at this early phase 
some field lines connected to the flaring region had opened, along 
which electrons could stream into the interplanetary space. 

About three minutes after the start of continuum emission, a metric type 
II burst appears in the dynamic spectrum, at 00:16:40 UT. The burst shows 
emission both at the fundamental ($\approx$\,120 MHz) and second harmonic 
($\approx$\,240 MHz) plasma frequencies. RSTN observations show that the 
emission lanes are band-split.  
Type II burst lanes are again observed at about 00:24 UT, now with  
emission at 50 MHz and just above 100 MHz. The emission bands of these 
bursts are outlined with dashed lines in Figure \ref{fig:fig4}. 
In between these two burst lanes, we see a pair of bursts that look like 
disturbed type III emission. The lower emission patch is located around
40 MHz at 00:21 UT. There have been reports on disturbed type III burst 
lanes, possibly formed when outward-streaming electrons travel through a 
turbulent shock region \cite{reiner00, lehtinen08}. So in our case the 
bursts probably trace the path of a propagating shock, when the shock 
no longer produces a continuous type II burst lane. 

The frequency drift of the first visible type II burst is approximately 
\mbox{0.4 MHz s$^{-1}$}. The drift rate of the second type II burst is about 
0.1 MHz s$^{-1}$. Both values are within the usual 0.1\,--\,1.0 
MHz s$^{-1}$ drift rate for typical type II bursts \cite{nelson85}.
No interplanetary type II bursts were observed associated with this 
event and no continuations of the metric type II bursts are visible 
in the dynamic spectra below 14 MHz ({\it Wind} WAVES observations).

\section{Analysis of the Radio Emission}

As plasma emission at the fundamental emission frequency  $f_p$ 
is directly related to electron density $n_e$, we can use radio observations 
to estimate shock heights and speeds under the assumption of a coronal 
density model. One should be aware that height estimates depend strongly on 
the density models used. By calculating heights with more than one model and 
by taking into account the local coronal conditions, it is possible to give 
reliable height ranges for the emission sources.  
Different models and their characteristics are explained in detail in, 
for example, \inlinecite{mann03}, Appendix A of  \inlinecite{vrsnak04}, 
and \inlinecite{pohjolainen06}. Direct observations of the densities 
of the pre-shocked coronal plasma are still rare, ({\it e.g.}, 
Mancuso, \citeyear{mancuso07}). 

First, we calculate burst heights with three different atmospheric 
density models, which describe different types of atmospheres. 
The basic (1$\times$) Saito (\citeyear{saito70}) model is often used 
for equatorial regions in the corona, and also at large distances from 
the Sun, where active region densities no longer dominate. 
A two-times Newkirk (\citeyear{newkirk61}) density model is often used 
for densities in streamer regions and in active region solar corona. 
A ten-times Saito model can be used when a shock is propagating through 
high-density loops in the low corona. 
Since the LASCO images show a streamer region in the northwestern part 
of the Sun where the CME appears (Figure \ref{fig:fig3}), high-density 
models should be more applicable. The estimated source heights at the 
start of the two bursts, and the derived velocities along the burst 
lanes, are given in \mbox{Table \ref{table1}}.

\begin{table}[h!]
\caption{Estimated heliocentric source heights and radio burst velocities}
\label{table1}
\begin{tabular}{ccccccc}
\hline
Time     &$f_p$ & $n_e$     & Drift rate  & $h$   & $h$            & $h$ \\
UT       &MHz   & cm$^{-3}$ & MHz s$^{-1}$ &Saito & 2$\times$Newkirk &10$\times$Saito \\
\hline
00:16:40 & 120 & 1.77$\times$10$^8$ & 0.4 & 1.09 $R_{\odot}$ & 1.30 $R_{\odot}$ & 1.46 $R_{\odot}$ \\
\multicolumn{4}{l}{Velocity}              & 1100 km s$^{-1}$ & 1700 km s$^{-1}$ & 2300 km s$^{-1}$  \\
00:24:00 &  50 & 3.08$\times$10$^7$ & 0.1 & 1.34 $R_{\odot}$ & 1.68 $R_{\odot}$ & 2.0 $R_{\odot}$ \\
\multicolumn{4}{l}{Velocity}              & 700 km s$^{-1}$  & 1000 km s$^{-1}$ & 1200 km s$^{-1}$ \\
\hline
\end{tabular} \\
\end{table}

We note that the radio source height given by the basic Saito model 
at 00:16:40 UT, is very near the projected EUV ejecta height at 00:16:50 UT 
(0.095 $R_{\odot}$ from the center of eruption). However, radio imaging 
observations at 164 MHz and frequencies above it usually show type II 
features near heights of 0.3\,--\,0.4 $R_{\odot}$ above the limb 
\cite{klassen99,dauphin06}. There are on-the-disk observations 
\cite{klassen03} of smaller projected distances from the flaring region, 
$\approx$\,0.2 $R_{\odot}$, but compared to these the height of 
$\approx$\,0.1 $R_{\odot}$ given by the basic Saito model for a source 
emitting at 120 MHz is uncommonly low. 
The heights given by the ten-times Saito model are also unrealistic at
these frequencies, since the source heights are in conflict with the 
densities observed above active regions (see, {\it e.g.} Figure 2 in  
\inlinecite{warmuth05}).
 
The values in Table \ref{table1} suggest that the two type II bursts 
were propagating at different speeds. However, a change in the coronal 
conditions
in between the bursts may change the speeds as well. For example, if the 
burst source first propagated in a low-density (Saito) atmosphere in the 
low corona and then entered a high-density streamer (2$\times$Newkirk) 
region, the shock speed would basically remain the same 
($\approx$\,1\,000 km s$^{-1}$), see the values in Table \ref{table1}. 
Transitions between different regions are possible since the shock can 
be spatially large, but in this case a low-density (basic) Saito model 
atmosphere very low above the active region is not probable.

Type II burst velocities can also be estimated with the scale
height method, see, {\it e.g.}, \inlinecite{demoulin00}. Shock 
velocity $v$ can be calculated from the observed plasma frequency $f_p$ 
(MHz), the measured frequency drift $df/dt$ (MHz s$^{-1}$), and the 
scale height $H$ (km) as 
  \begin{equation}
   v = 2 \frac{1}{f} \frac{df}{dt} H .
  \end{equation}
\inlinecite{koutchmy94} discovered that the hydrostatic isothermal 
scale height in a \mbox{1.5 MK} coronal temperature describes well 
the measured density profiles in the solar atmosphere, regardless of 
the base density. 
Since the source heights presented in Table \ref{table1} and the 
discussion on the values suggest that heights near 0.2\,--\,0.4 $R_{\odot}$ 
above the photosphere are possible for a radio source emitting
at 120 MHz, we calculate the speeds using the scale height method. 
At a heliocentric height 1.2 $R_{\odot}$ the scale height is then 109\,000 km. 
The emission frequency at the start of the first type II burst was 120 MHz 
and the observed frequency drift was 0.4 MHz s$^{-1}$,  which gives a speed of 
725 km s$^{-1}$. If we assume that the radio source was located higher, 
at 1.4 $R_{\odot}$, the scale height is 148\,000 km and the corresponding 
speed is 985 km s$^{-1}$. If the shock propagated at 725 km s$^{-1}$ 
from a height 1.2 $R_{\odot}$, it would have reached the height of 
1.66 $R_{\odot}$ at the time when the second type II burst started. But, 
the burst velocity calculated from the emission frequency of 50 MHz and 
the frequency drift of 0.1 MHz s$^{-1}$ at that height does not match 
with the first type II burst (being in fact higher). 
If, however, the second type II burst originated from a lower height, 
at 1.55 $R_{\odot}$, the calculated shock speeds for the first and second 
type II bursts would be the same.
This implies that a deceleration of the shock would not be able to
produce the observed frequency drift of the second type II burst, 
but a driver propagating at a constant speed and producing shocks at 
different parts (heights) of the expanding body could explain the two 
separate type II bursts.
               
\begin{figure}
\includegraphics[width=10cm,angle=0]{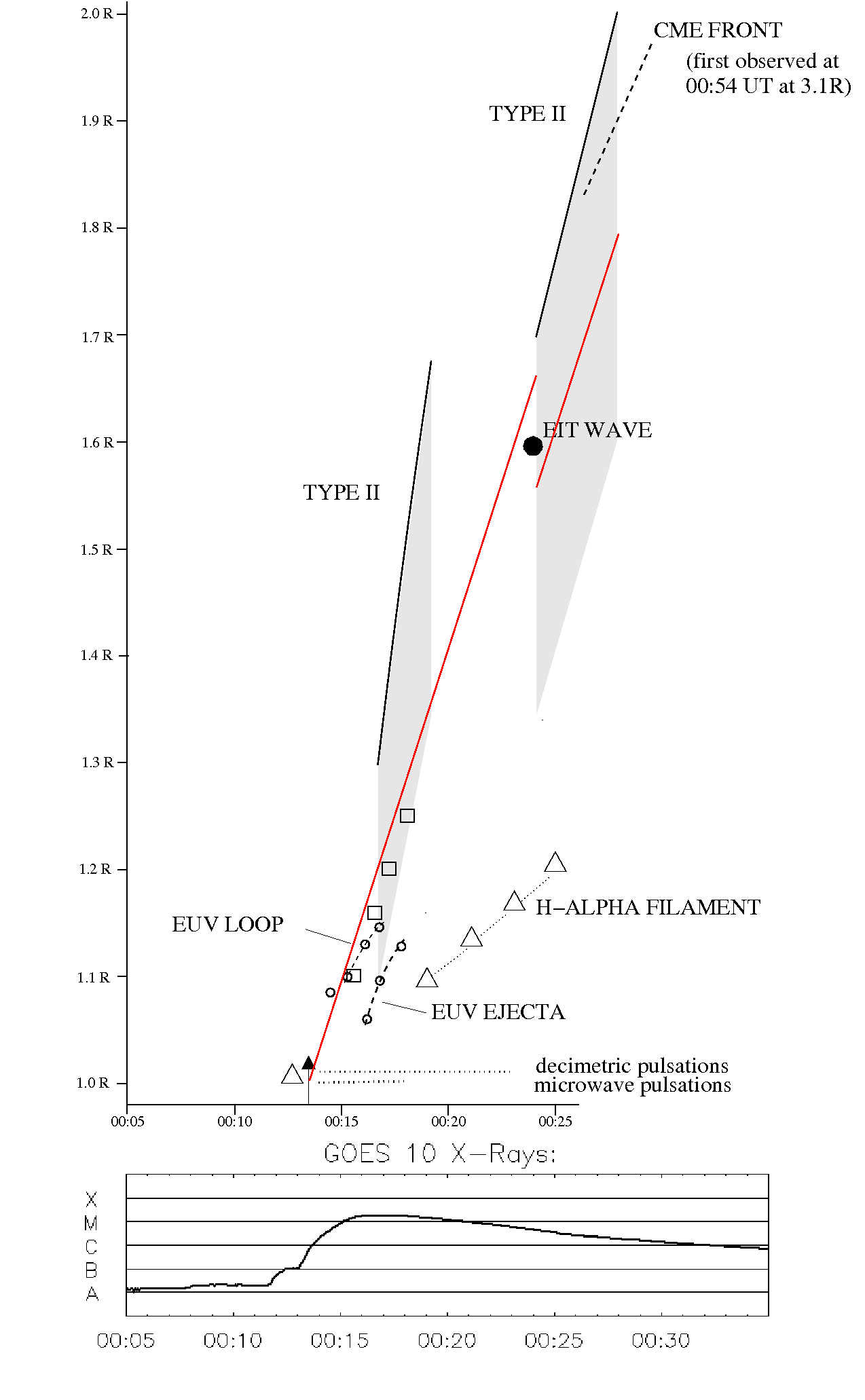}
  \caption{Estimated (projected) distances of the rising EUV loop and 
  ejecta (circles), H$\alpha$ Moreton wave (boxes), EIT wave (visible 
  at 00:24 UT, but no wave detected in the previous image at 00:12 UT), 
  backward-extrapolated CME front locations, and estimated heights for 
  the type II bursts using a 2 $\times$ Newkirk density model (solid black 
  lines). The shaded areas for the type IIs represent height values that 
  can be obtained with less-dense atmospheric density models. The lower 
  limits are when using the basic Saito density model (see text for details).
  Red lines give the height--time trajectories for a disturbance 
  traveling at a speed of 725 km s$^{-1}$ and appearing at a height 
  of 1.2  $R_{\odot}$ at 00:16:40 UT and at a height of 1.55 $R_{\odot}$ 
  at 00:24 UT (calculated from the frequency drifts and scale heights).    
  The time periods for the microwave and decimetric pulsations are 
  indicated with dotted lines, and black arrow marks the start time
  of the drifting continuum emission near 1 GHz.
  Triangles indicate the start time of the filament eruption and the  
  estimated later heights obtained from the H$\alpha$ observations. 
  The GOES soft X-ray flux curve at 3\,--\,25 keV is shown at the bottom. } 
  \label{fig:fig5}
\end{figure}

Figure \ref{fig:fig5} summarizes the height versus time relations for
the metric type II bursts. Height ranges are shown from the basic
Saito and two-times Newkirk atmospheric models.  Figure \ref{fig:fig5} 
also shows the height--time trajectory for a disturbance traveling at 
725 km s$^{-1}$, passing the height of 1.2 $R_{\odot}$ at the time of the 
first type II burst appearance. A disturbance propagating at the same
speed would have to be located at a lower height at the time of the
second type II burst.
The projected distances for the EUV loop front, EUV ejecta, Moreton wave, 
apparent front of the ejected H$\alpha$ filament, EIT wave, and the 
backward-extrapolated CME front are also given. In the plot, the time 
periods of radio pulsations, start time of the drifting continuum emission, 
and the GOES soft X-ray flux are shown for reference.

\begin{figure}
\includegraphics[width=12cm,angle=0]{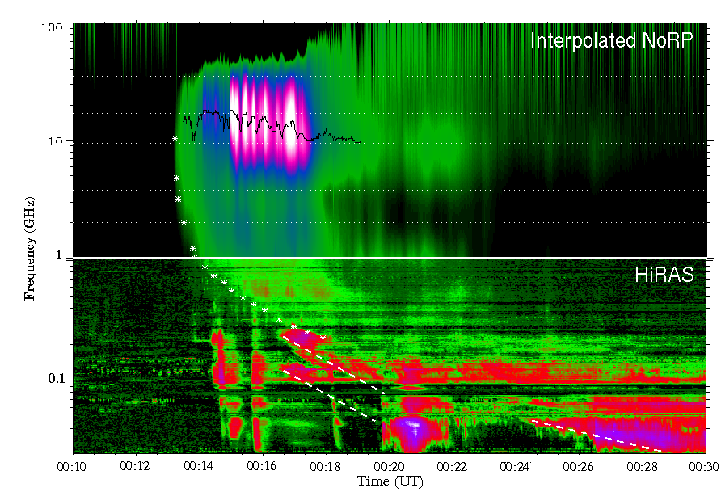}
 \caption{Interpolation of NoRP single-frequency observations above
1 GHz (with the dotted horizontal lines showing the observing frequencies) 
combined with the HiRAS dynamic spectral data. The plot shows a 
possible expansion front (marked with stars), which can be associated with 
the CME and the type II bursts. The first type II burst (with fundamental and
second harmonic emission lanes indicated with dashed lines) appears in a 
region with lower electron density/lower plasma frequency), and the
later second type II burst (with only one emission lane marked here) falls
approximately within the drifting continuum. Emission at microwaves 
show quasi-periodic pulsations (not discussed in this paper).} 
\label{fig:fig6}
\end{figure}

To analyze the drifting continuum emission we cannot calculate  
height estimates, since the frequency drifts can indicate both expanding 
volume (decreasing density within the volume) and/or rising structures 
(decreasing atmospheric density). 
We compared the HiRAS dynamic spectrum with the six NoRP single-frequency 
observations at 1, 2, 3.75, 9.4, 17, and 35 GHz. We interpolated the radio 
intensity over 1\,--\,100 GHz using a least squares fit for the six 
frequencies. The interpolated data suggest a possible 
expansion front -- defined here as the leading front or border of a 
fast-expanding structure -- and visible as the leading edge of 
the continuum emission (Figure \ref{fig:fig6}). 
This leading edge is marked with stars in the plot. 
Electrons may have been trapped in, for example, an expanding solar magnetic 
arch and emitted gyrosynchrotron radiation. The continuum could then reveal 
radiation from nonthermal electrons confined within the plasma volume, 
which drives a shock. An expansion front would be located at the
outer boundary of the expanding volume, and the shock would be located 
ahead of it. The distance between the driver and the shock is discussed
more in Section 4. 

The continuum appeared at about the time of the filament eruption, at  
00:13 UT, which is almost two minutes before eruption or shock signatures 
at other wavelengths were observed. The computed kinematics presented by 
\inlinecite{grechnev08} for the Moreton and EIT waves, also suggest that 
the waves were formed later, around 00:15 UT.

\section{Results and Discussion}

We can summarize the multi-wavelength analysis of the 13 July 2004  
event with the following results:

1. The event started with a filament eruption, followed closely by 
the appearance of a frequency-drifting radio continuum. The constructed 
microwave--meterwave dynamic spectrum (Figure \ref{fig:fig6}) suggests 
that an expansion front was formed at the time of the filament eruption, 
almost two minutes prior to any signatures of shock wave formation at
other wavelengths. This expanding plasma volume could have been driving 
a shock wave.

2. The first type II burst was observed in the dynamic spectrum soon 
after the EUV ejecta lifted off (from below the EUV loop system) and 
the Moreton wave appeared. 
The estimated type II burst velocity is similar to the Moreton 
wave velocity ($\approx$\,700 km s$^{-1}$), if the type II burst source 
height is set untypically low, \mbox{$\lesssim$ 0.2 $R_{\odot}$} above
the photosphere. The type II burst heights then match with the projected 
distances of the Moreton wave front. 
We do not expect the Moreton wave and type II burst speeds to be exactly 
the same. According to Uchida's model \cite{uchida74}, the H$\alpha$ 
Moreton wave is created when the "skirt" of a shock wave sweeps the 
chromosphere. The shock velocity upward probably differs from the 
horizontal velocity, but the observed velocities can still be used as 
rough estimates for the shock speed (like the type II speeds). The 
backward-extrapolation of a shock propagating with a constant speed of 
725 km s$^{-1}$ suggests a common start time with the frequency-drifting 
radio continuum (Figure \ref{fig:fig5}).
For comparison, at more typical type II heights, 
$\approx$\,0.3\,--\,\mbox{0.4 $R_{\odot}$}, the calculated burst velocities 
get much higher, $\approx$\,1\,000\,--\,\mbox{1\,700 km s$^{-1}$}, and 
no correlation with  the heights of other features can be found.      

3. The second type II burst had a frequency drift that suggests  
a speed different from that of the first type II burst, thus implying a 
different driver for the shock. Calculations show that the type II bursts 
could have been formed by one shock only, if the shock propagated 
at a constant speed and the later shocked region was located lower 
in the shock driver structure ({\it e.g.}, if the first type II burst was 
created at the nose of expanding loops and the second one on the sides). 
The estimated height separation, \mbox{$\approx$\,0.1 $R_{\odot}$}, is 
not large enough to indicate a separation between, for example, a shock 
in front of a CME and at the flanks. A decelerating shock would not have 
been able to produce the observed frequency drift of the second type II burst. 
A single shock propagating through regions with very different 
densities, such as first through low density structures and later 
through high-density streamer regions, would also produce similar 
jumps in the type II source heights. 
Despite these different interpretations, the estimated heights for the 
second type II burst are close to the observed EIT wave distance and 
the backward-extrapolated CME front distances, although the speeds 
differ considerably.
   
Freely propagating shock waves can be formed in flare blasts or when 
a piston is stopped or it slows down ({\it e.g.}, \opencite{vrsnak01}). 
The speed of the propagating shock and the speed of the initial driver 
can then differ considerably. Also the shock should be spatially well 
separated from the driver. (In CME-driven shocks observed near Earth, 
the standoff distance between the shock and its driver has grown to 
20\,--\,50 R$_{\odot}$, see Reiner {\it et al.}, \citeyear{reiner07}). 
\inlinecite{pomoell08} have shown in their simulations how the speed of 
the shock (and hence the corresponding type II burst speed) can far exceed 
the speed of the driver, and the shock can continue propagation through 
the corona even if the driver speed falls below the local magnetosonic speed.
Recent radio imaging observations \cite{dauphin06} show a type II 
burst located ahead of a rising soft X-ray loop. The speed of the rising 
loop was only 650 km s$^{-1}$ at the start time of the type II burst. 
As stated in their paper, and in another paper analyzing the same event 
\cite{vrsnak06}, the derived speeds for the type II shocks   
(obtained with different methods, but in the range of 1\,100\,--\,1\,800 
km s$^{-1}$) are much higher than those of their assumed drivers.

Based on the theory and our calculations for burst heights and 
speeds, and the observation of the Moreton wave, it is possible that 
the first type II burst was formed by a blast wave. The earlier start 
of the frequency-drifting radio continuum suggests that a shock wave 
may also have been ignited by a fast-expanding structure.
Loop-like fronts, such as the rising EUV loop in this case, can be the 
low-coronal signatures of CMEs \cite{rust76,vrsnak04b,temmer08}, that 
drive the shock. \inlinecite{gary84} presented a case where
a shock is driven by rising loops and a type II burst is due to a 
flare blast that starts later. In this scenario the type II burst
source is located within the ejecta material. Type II emission can 
also be produced when a flare shock overtakes regions of principal
density pile-up near the base and sides of an expanding transient 
\cite{wagner83}. This could also explain our low radio source heights.

\inlinecite{subra06} have analyzed statistically several doublet type II 
bursts, and concluded that usually the second type II burst starts at a 
lower frequency than the first one and the drift rate is only about half 
of the first one. In our event the frequency drift of the second type 
II burst is slower by a factor of 4. The rare imaging observations of 
multiple type II bursts \cite{robinson82} suggest that a single shock 
may exist, but as the shock propagates at different locations the type 
II bursts may have different characteristics. To verify this,
imaging observations are needed at radio and other wavelengths, to 
determine source locations and corresponding electron densities.

We have presented analysis of a very complex flare-CME event, with a 
multitude of radio features and transients at other wavelengths, and 
obtained evidence for both blast wave shock and CME-related emissions 
during the initial phase. However, definite conclusions will rely on 
future radio imaging. It is unfortunate that radio imaging currently exists  
only at very narrow frequency ranges, leaving out plasma emission above 
500 MHz and below 150 MHz. For the identification of the source locations
in decimetric--metric radio emission, future radio imaging instruments 
such as the Frequency-Agile Solar Radiotelescope (FASR) will be needed,
together with high-cadence multi-wavelength observations. Statistical 
analyses with comparison to eruption characteristics could also clarify 
the processes involved.

\begin{acknowledgements}
We wish to thank K-L. Klein for discussions on the use of coronal
density models and Victor Grechnev for additional information and 
discussions on the analysed event. The anonymous referees and Guest 
Editor K-L. Klein are thanked for valuable comments and suggestions 
on how to improve the paper.    
The Global High Resolution H$\alpha$ Network is operated by the Big
Bear Solar Observatory, New Jersey Institute of Technology, and we
are grateful to V. Yurchyshyn for providing the H$\alpha$ data.
We have used in this study radio observations obtained from the 
Hiraiso Solar Observatory (National Institute of Information and 
Communications Technology) and Nobeyama Solar Radio Obsevatory (National
Astronomical Observatory of Japan), and we thank their staff for making 
the data available. {\it Wind} WAVES data can be obtained from their 
web archive. 
SOHO is a project of international cooperation between ESA and NASA.
The SOHO LASCO CME Catalog is generated and maintained at the CDAW 
Data Center by NASA and the Catholic University of America in cooperation 
with the Naval Research Laboratory.  
We are grateful to the TRACE, SOHO EIT, LASCO, and MDI teams for making
the data available at their Web archives.
S.P. and K.H. acknowledge travel grants for this work from the Academy 
of Finland Project No. 104329.
\end{acknowledgements}

\end{article}

\end{document}